\documentclass{aa}
\usepackage{natbib}
\usepackage{longtable}
\usepackage{lscape}
\bibpunct{(}{)}{;}{a}{}{,}
\usepackage{graphicx}
\usepackage{txfonts}
\usepackage{xcolor}

\begin{document} 

   \title{Wide binaries in Planetary Nebulae with Gaia DR2} 
   
   \titlerunning{Binary stars in PNe with Gaia DR2}

   \author{I. Gonz\'alez-Santamar\'{\i}a\inst{1,2} \and M. Manteiga\inst{2,3} \and A. Manchado\inst{4,5,6} \and M.A. G\'omez-Mu\~noz\inst{4,5,7} \and  A. Ulla\inst{8} \and C. Dafonte\inst{1,2}}
   \authorrunning{I. Gonz\'alez-Santamar\'{\i}a et al.}
   \institute{Universidade da Coru\~na (UDC), Department of Computer Science and Information Technologies, Campus Elvi\~na s/n, 15071 A Coru\~na, Spain \\
  \email{iker.gongalez@udc.es}
  \and
  CITIC, Centre for Information and Communications Technologies Research, Universidade da Coru\~na, Campus de Elvi\~na s/n, 15071 A Coru\~na, Spain                    
  \and
   Universidade da Coru\~na (UDC), Department of Nautical Sciences and Marine Engineering, Paseo de Ronda 51, 15011, A Coru\~na, Spain \\
   \email{manteiga@udc.es}
  \and
  Instituto de Astrofísica de Canarias, 38200 La Laguna, Tenerife, Spain
  \and
  Universidad de La Laguna (ULL), Astrophysics Department
  \and 
  CSIC, Spain
          \and
   Instituto de Astronom{\'i}a, Universidad Nacional Aut{\'o}noma de M{\'e}xico, Apdo. Postal 877, 22800 Ensenada, B. C., Mexico          \and    
  Universidade de Vigo (UVIGO), Applied Physics Department, Campus Lagoas-Marcosende, s/n, 36310 Vigo, Spain
  \\
           }
   \subtitle{}

   \date{Received 14-09-2020/Accepted 09-11-2020}
 
  \abstract
   {Gaia Data Release 2 (DR2) 
   was used to select a sample of 211 central stars of planetary nebulae (CSPNe) with good quality astrometric measurements, that we refer to as GAPN, Golden Astrometry Planetary Nebulae. Gaia astrometric and photometric measurements allowed us to derive accurate distances and radii, and to calculate luminosities with the addition of self-consistent literature values. Such information was used to plot the position of these stars in a Hertzsprung-Russel (HR) diagram and to study their evolutionary status in comparison with CSPNe evolutionary tracks.}
   {The extremely precise measurement of parallaxes and proper motions in Gaia DR2 has allowed us to search for wide binary companions in a region close to each of the central stars in the GAPN sample. We are interested in establishing the presence of binary companions at large separations which could allow to add further information on the influence of binarity on the formation and evolution of planetary nebulae. We aim to study the 
   evolutive properties of the binary pairs to check the consistency of spectral types and masses in order to better constrain the ages and evolutionary stage of the CSPNe.} 
   {We limited our search to a region around 20,000 AU of each CSPN to minimise accidental detections, and only considered stars with good parallax and proper motions data, i.e. with errors below $30\%$ in DR2. We determined that the hypothetical binary pairs should show a statistically significant agreement for the three astrometric quantities, 
   i.e. parallax and both components of proper motions.}
   {We found 8 wide binary systems among our GAPN sample. One of them is in a triple system. We compiled the astrometric and photometric measurements of these binary systems and discussed them in comparison with previously published searches for binaries in PNe. By analysing the 
   position in the HR diagram of the companion stars using Gaia photometry, we are able to estimate their temperatures, luminosities, masses and, for one star, the evolutionary age. The derived quantities yield a consistent scenario when compared with the corresponding values as obtained for the central stars using stellar evolutionary models in the postAGB phase.}
   {}

\maketitle


\section{Introduction}
In a recent work, \citet{2019A&A...630A.150G}, which we will refer to as paper 1, we used Gaia Data Release 2 (DR2) astrometry to build a catalogue with central stars of planetary nebulae (CSPNe) precise distances and luminosities, that we named Golden Astrometry Planetary Nebulae (GAPN). Gaia DR2 information was analysed together with literature values for the extinction, stellar temperatures, expansion velocities and nebular radii to estimate other parameters such as physical sizes and kinematical ages. We were able to locate the CSPNe in a Hertzsprung-Russel (HR) diagram and to compare their distribution with that predicted by up-to-date evolutionary tracks. GAPN includes 211 CSPNe, which were selected by applying quality filters to the astrometric measurements (parallax relative errors, as well as distances derived errors, lower than 30{\%}).

Several previous works have taken advantage of the extraordinary quality of Gaia DR2 astrometry to search for wide binaries among the brightest Milky Way stars \citep{2019AJ....157...78J}, and have proven that it is possible to search for comoving objects by considering proper motions and parallaxes (and radial velocities, if they are available) based on Gaia or a combination of Gaia and Hipparcos data. In this work we aim to carry out a systematic search for comoving systems in the field around each nebular central star (CS) in the GAPN sample. This way, we intend to shed more light on the phenomenon of binarity in planetary nebulae, adding to the search for binary companions close to the central stars, the possible existence of more separated
objects, with likely a less direct influence on the evolution of the CS, but equally interesting when it comes to constraining values such as the age and mass of the CS. Previous studies of wide binary companions to planetary nebulae (PNe) (in particular the HST snapshot program from \citealt{Ciardullo_1999}) were successful in providing accurate distances to PNe by measuring the photometry of nearby objects that were postulated as probable binary partners. Gaia DR2 measurements will allow now to confirm some of such derivations.

A recent review by \citet{2019ibfe.book.....B} analyses the importance of binaries in the formation and evolution of PNe, insisting on the idea that most of them could be ejected from binaries but stressing the fact that providing a quantitative value about the incidence of binarity is not an easy task. Nevertheless, binarity is one of the most accepted theories to explain asphericity in the nebular morphology, being aspherical approximately 75\% of the total of known PNe  \citep{2004ASPC..313....3M,2017NatAs...1E.117J}. 
In addition, in \citet{2001ApJ...558..157S}, some morphological properties, such as departure from axisymmetry and the formation of bipolar and multipolar lobes that support the binary model, are discussed. 

In the case of close binaries, with orbital separations of the order of a few AU, there can be a strong interaction between both stars. The interaction could involve the spin-up of the envelope by tidal forces, Roche lobe overflow, and common envelope (CE) evolution.
In this phase, mass transfer between both stars is produced either by stellar winds or by Roche-lobe overflow. According to \citet{2020Galax...8...28J} approximately 12-20 \% of PNe host a post CE system, and in such systems it has been hypothesised that the ingestion of the companion can lead to the bipolar ejection of the progenitor envelope, strongly affecting the PN morphology. 
Other authors propose that more distant binary companions may provide an alternative mechanism for the formation of highly aspherical morphologies by influencing the direction of collimated winds from the parent star, whose axis of rotation precesses due to the influence of a companion that is not necessarily close (\citealt{GarciaSegura97}, see discussion in \citealt{1998AJ....116.1357S}).

In wide binaries, the link is not so strong, but knowing some properties - such as the common age of the components or the orbital separation - can give certain clues on the evolution of the CSPNe and also on the possible influence of wide companions in the nebular morphology. \citet{2017ApJ...837L..10B} suggested that triple systems could shape the morphology of those PNe that depart from axial-symmetry or mirror-symmetry, and hierarchical systems have been found in some PNe (for instance in NGC 246, \citealt{10.1093/mnras/stu1677}). In our work we are focusing on wide binaries, that can be detected as separated sources in Gaia DR2. Future Gaia archives, which will include the analysis of the astrometric solution for several tens of observational epochs, will hopefully lead to finding some astrometric binary companions at angular distances as short as 0.1 arcsecs among the brightest CSPNe.

The present paper is structured as follows: in section 2, we describe the methodology chosen to search for comoving objects, which is partially based on the procedure presented by \citet{2019AJ....157...78J}, and we discuss the reliability of the possible binary partners, to finally propose our list of solid binary candidates.  
 In section 3, we present the astrometric and photometric properties of the binary pairs detected. We used Gaia DR2 three bands photometry together with literature information about the interstellar extinctions towards the systems, to locate
 the companions of the CSPNe in a HR diagram. When Gaia colours were not available, we used VOSA tool from the Spanish Virtual Observatory \citep{2008A&A...492..277B} to construct the spectral energy distribution from the available literature spectrophotometric data.   
 Evolutionary models could then be fitted to obtain effective temperatures, luminosities and masses (and their uncertainty values), to be compared with the ones inferred for the CSPNe in the GAPN sample. Section 4 discusses the results obtained, which can clearly be expanded upon when the third Gaia archive will be published in 2021.


\section{Searching for comoving binary pairs}


In order to search for comoving objects within our GAPN sample, we have to analyse the density of stars and their astrometric properties in a field close to each nebula. 
We decided to limit the study of fields to stars with magnitudes G lower than 19, as weaker objects usually have significant errors in their DR2 astrometric measurements. Comoving objects are those that have values of their five astrometric parameters (position in the sky, parallax and proper motions in right ascension, and declination) plus radial velocities, consistent with spatial closeness and gravitational linkage between them. In this case, radial velocities are not available for the companion candidates and we had to constrain our search to compatibility of the other five parameters. 

To ensure a physical relationship, it is important to exclude the possibility of similar astrometry occurrence by chance, as well as to take into account the errors of the astrometric measurements in DR2. In what follows we describe how we proceeded. We retrieved the astrometric measurements and their errors for all stars brighter than G = 19 in a 120 arcsec-wide region around the coordinates of each of the GAPN CSs. This angular radius was chosen as a balance between achieving sufficient statistics to measure the density of stars and their astrometric parameters, while covering a region relatively close to each nebulae (it corresponds to a projected radius of 0.58 pc for a nebula at a distance of 1 kpc).
On average, we found approximately 250 objects around each of the considered CSPNe. 



If we define $\sigma$ as the largest value of the error in parallax (or in proper motions) for a given pair of a field object and a particular CSPN, we selected those objects whose parallax and proper motions differ less than $2.5 \cdot \sigma$ from the CS ones, the same criteria as those used by \citet{2019AJ....157...78J}. Furthermore, in order to pick up objects with enough reliable astrometry, we only considered objects with parallax and proper motions relative errors smaller than 30\%. Note that both parallax and proper motions in Gaia DR2 need to be corrected (for a systematic zero point and rotation bias, respectively) and that to estimate the errors in the astrometric parameters both internal and systematic errors have to be considered as prescribed in \citet{lindegren18}. 


For each of the 211 sources in the GAPN sample, we looked for companions with compatible parallax and proper motions values within errors ($2.5 \cdot \sigma$, as mentioned).
After applying these constraints, we found 44 objects distributed in 31 different systems, i.e. 31 out of 211 CSPNe have at least one object as a candidate to be a comoving companion. 

Physical separation, not angular separation, should be used as a general criterion to search for binary companions. In the case of wide binaries, generally accepted is a limiting maximum value of 20,000 AU in projected physical separation of binaries \citep{1987ApJ...312..367W, 1990AJ....100.1968C}, although this value is questioned in the case of sparse halo field stars or very young clusters (see \citealt{2019AJ....157...78J} for a discussion). A recent study by \citet{2020AJ....159...33Z} has proposed a value of 0.1 pc ($\sim$ 20,000 AU) for the maximum projected physical separation between binary components. Our 44 comoving candidates are located at distances from the CSPNe that range from 1.8 to 119.3 arcsecs, and projected physical separation distances ranging from 1,500 to 410,000 AU.

Planetary nebulae have physical sizes that vary between less than 0.1 pc in the earlier 14 kyr of their evolution \citep{2019A&A...630A.150G} to values of more than 1 pc when they are old diluted nebulae. Limiting the search of binary companions to comoving objects at projected separation distances not greater than 20,000 AU implies that, in general, we will be searching inside the nebular area. As a result of setting this limitation to our sample, only 8 out of the 31 systems remained as possible comoving systems (one of them as a triple system). Identification charts for the CSPNe and the comoving companions is provided in Fig. \ref{fig:images}.

\begin{figure}[h!]

        \includegraphics[width=4.46cm,height=4cm]{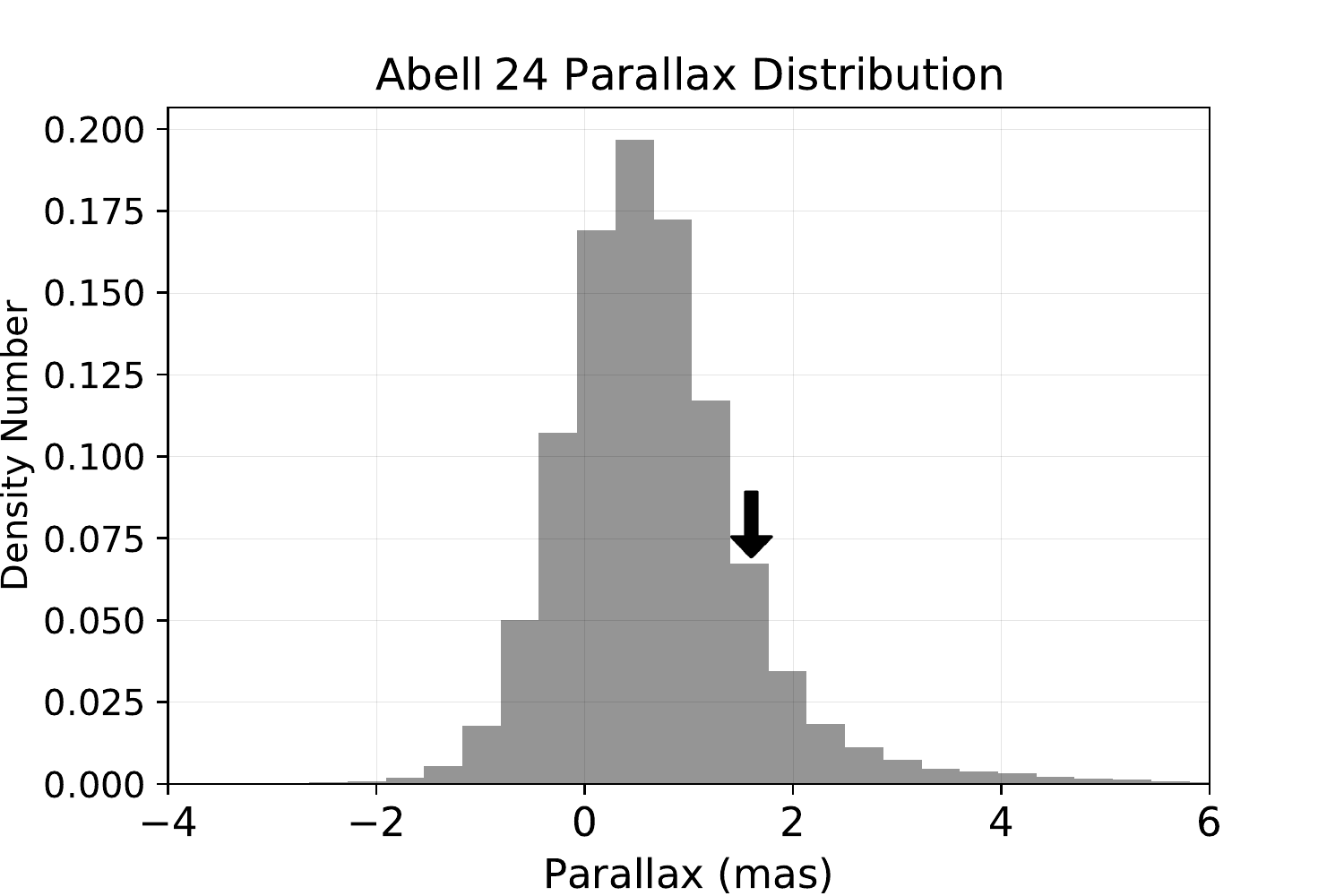}
        \includegraphics[width=4.46cm,height=4cm]{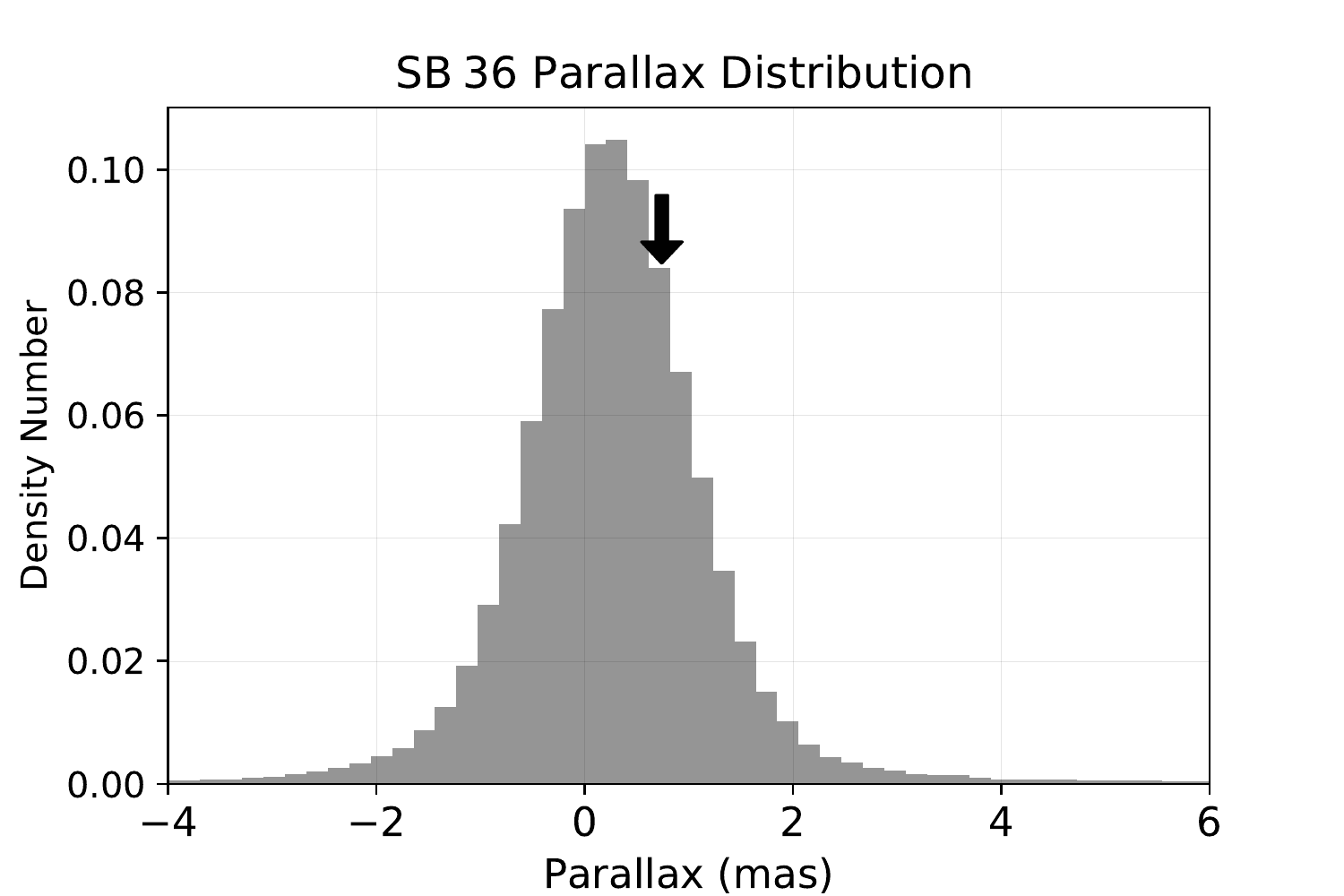}        
        \includegraphics[width=4.46cm,height=4cm]{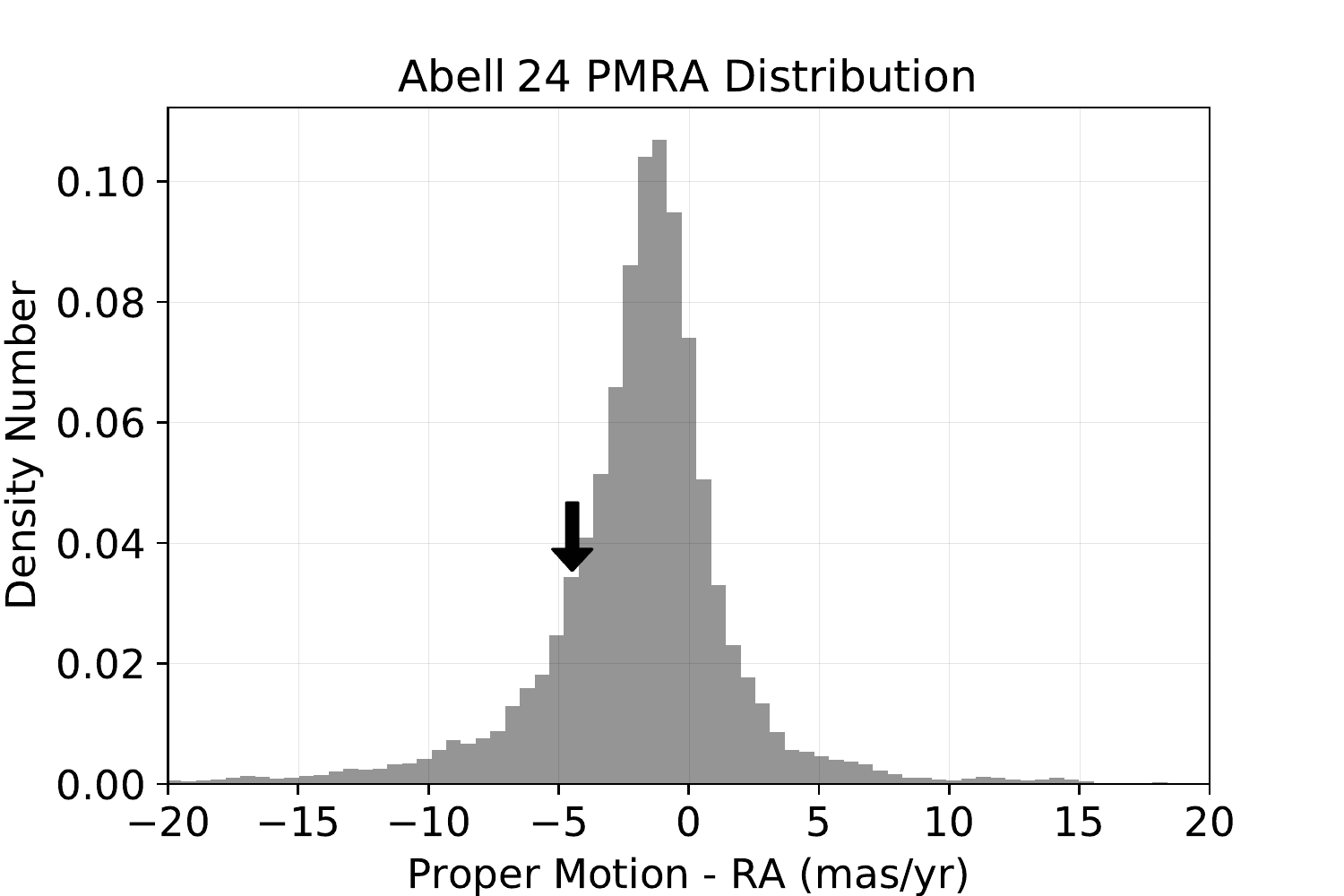}
        \includegraphics[width=4.46cm,height=4cm]{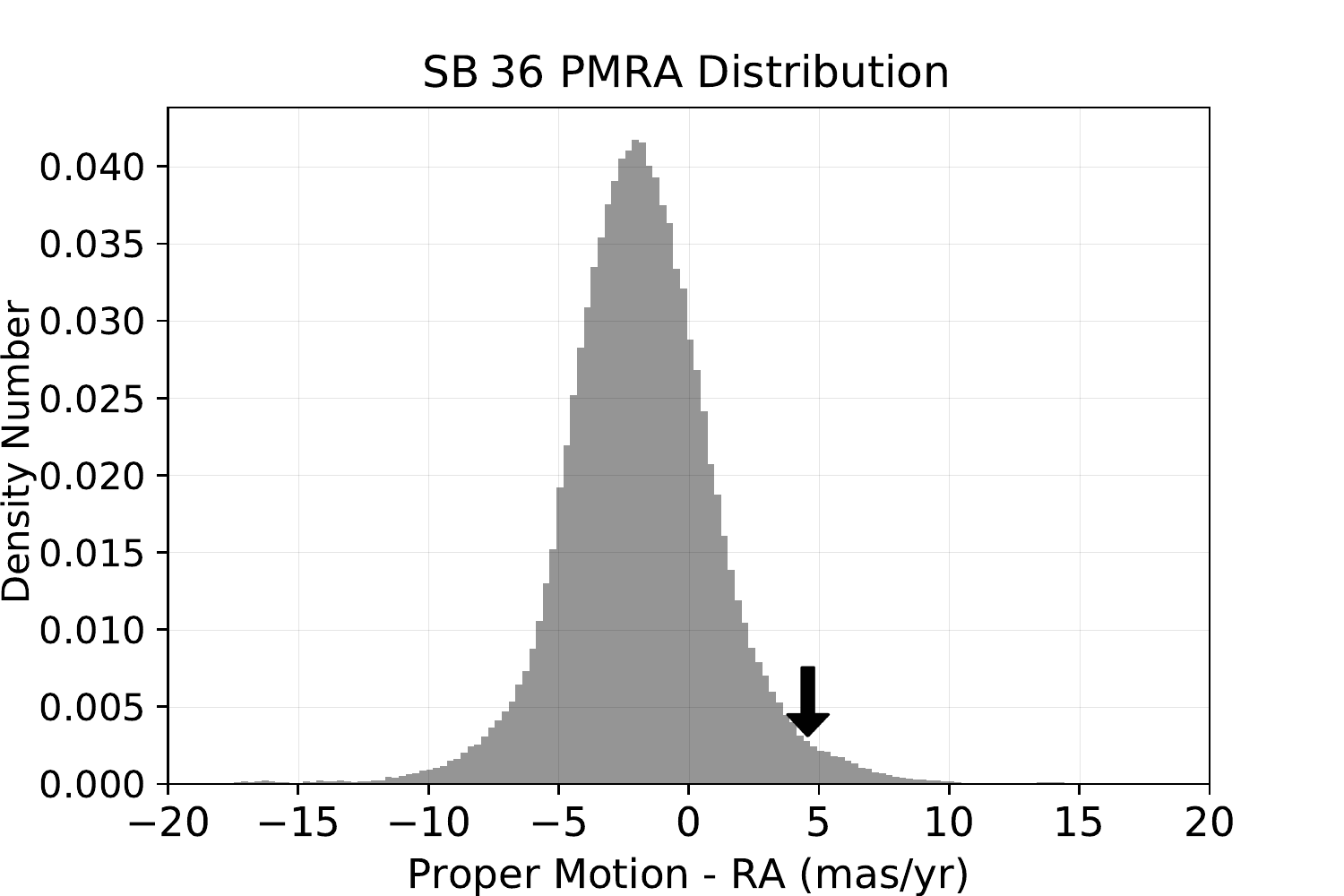}        
        \includegraphics[width=4.46cm,height=4cm]{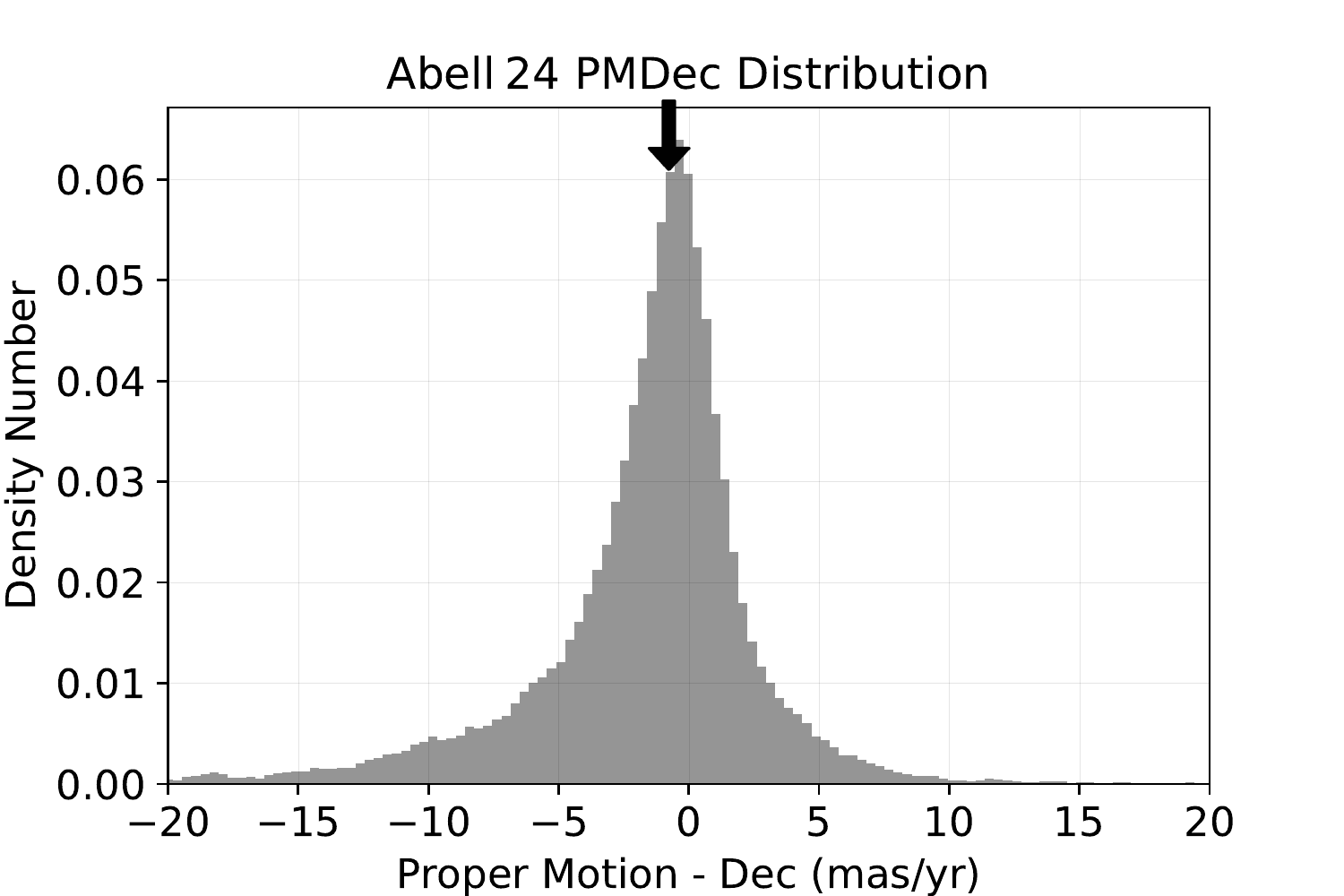}
        \includegraphics[width=4.46cm,height=4cm]{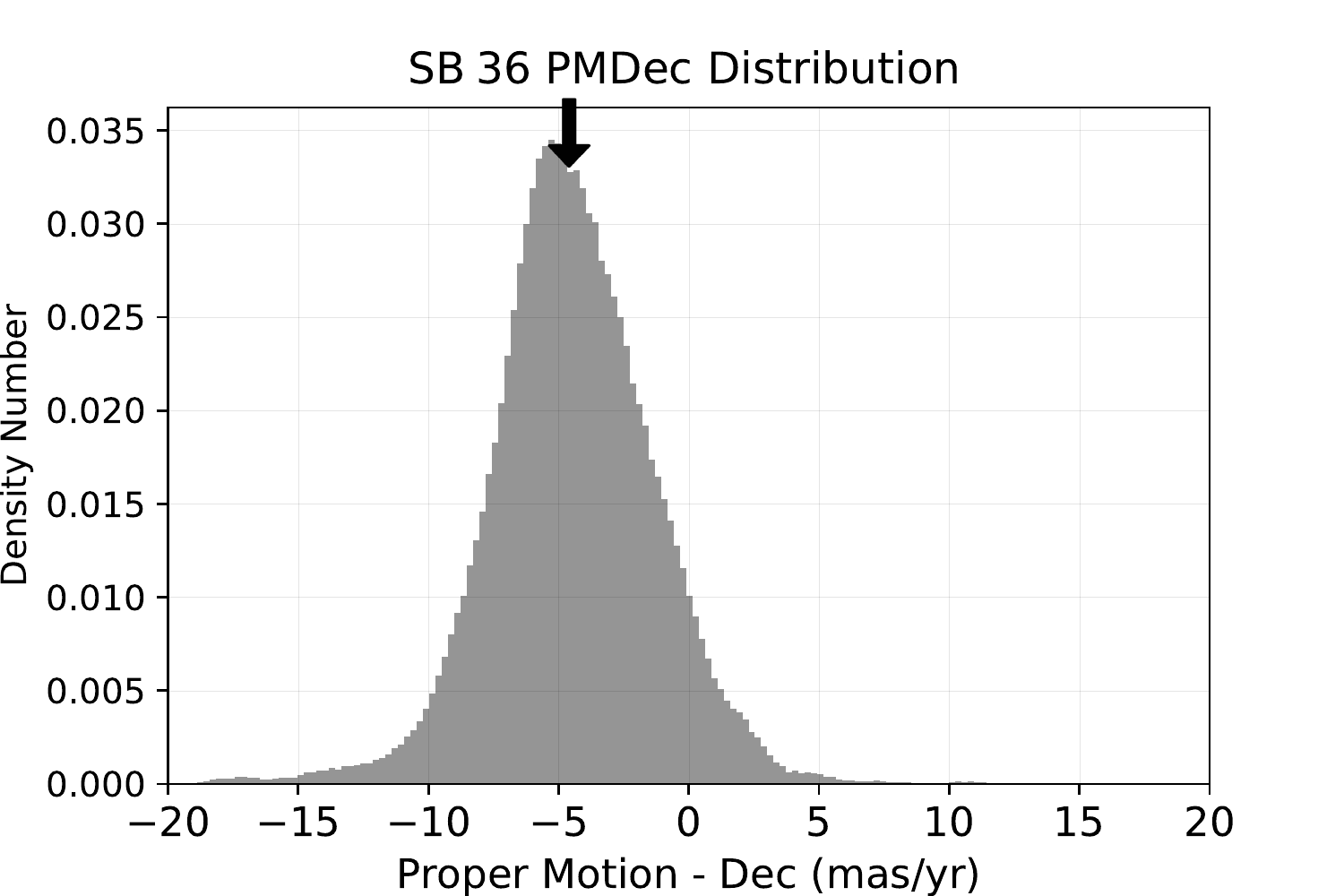}
        \caption{Probability distributions for parallax and proper motions of objects in the neighbourhood of Abell 24 (left) and PN SB 36 (right).The black arrow indicates the interval corresponding to the CS parameters values.}
        \label{fig:prob}
\end{figure}

We carried out a test to assess the degree of chance coincidences of the astrometric values for each of our 8 detected comoving pairs. We calculated the probability density distribution of the three astrometric measurements (parallax and proper motions) around each of the nebulae. In order to have enough statistics to draw the probability distribution, we considered spherical regions around the CSPNe with radii that vary from 1 to 4 pc depending on the obtained population, which was always above 400 objects. We limited our search to stars with magnitude G < 19, and we did not impose any constraint on the relative errors of the astrometric measurements. In fact, each of the astrometric quantities in DR2 is affected by errors that in some cases can get quite large, including negative parallaxes \citep{lindegren18}. 
A gaussian error distribution for every measurement was assumed, and we sampled every measurement 100 times in order to properly take into account the errors. Then, to build the distributions, we chose a sampling interval equal to $2.5 \cdot \sigma$, being $\sigma$ the error in the astrometric quantity previously used to search for comoving objects. This way, we were able to calculate the probability for every astrometric measurement in the region close to the CSPNe and, by combining the probabilities for the parallax and both proper motion parameters, we obtained a probability value for the specific triplet ($p, \mu_{\alpha}, \mu_{\delta}$). Such total probabilities always resulted very low, 
below $0.2\%$. 
To illustrate this procedure, Fig. \ref{fig:prob} shows the probability density distributions for parallax ($p$), proper motion in RA ($\mu_{\alpha}$) and proper motion in DEC ($\mu_{\delta}$) around Abell 24 and PN SB 36, two of the PNe with binary candidates.

Consequently, our search and selection procedure led us to propose a sample of 8 comoving systems that we interpret as wide binary systems associated to the CSPNe in GAPN. The rest of the present work is focused on that sample.


\section{Results}





Table \ref{tab:table1} contains the astrometric data for the 8 central stars and their detected binary companions. CSs distances are obtained from 
paper 1, while companion stars distances are obtained from Gaia DR2 archive by the same method used in that work. Other parameters, such as projected physical separations between binaries or nebular sizes, are derived from these distances.   
As can be seen, angular separations range from 1.8 to 13.7 arcsecs and projected separations are between 1,542 to 14,880 AU. As these are typical separation distances for wide binaries, they are expected to be large enough to discard the phenomenon of mass transfer. It should be noted that we have used a 30\% uncertainty threshold for the astrometry, the same  threshold that we used in \citet{2019A&A...630A.150G} where we defined the GAPN sample. Despite this, the relative errors of parallax in the measurements of the CS and its companions that are listed in Table 1 are always below 22\%, with only 4 measurements that exceed 15\%.
 
\begin{figure}[h!]
        \includegraphics[height=6cm,width=9.5cm]{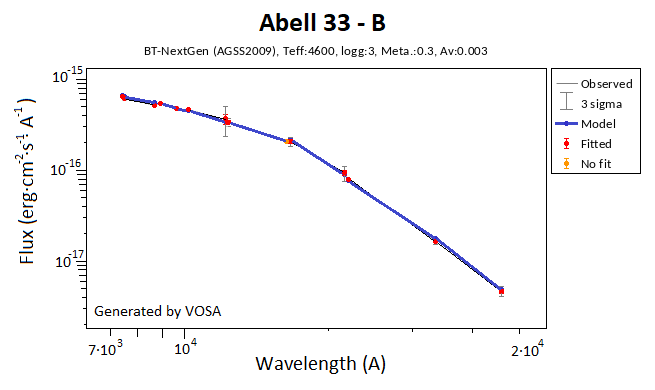}
        \includegraphics[height=6cm,width=9.5cm]{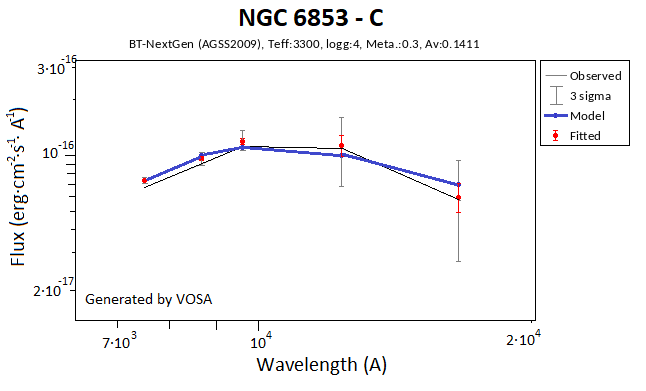}
        \caption{Fits to NextGen (AGSS2009) models of Abell 33 - B and NGC 6853 - C spectral energy distributions obtained with VOSA tool from the Spanish Virtual Observatory.}
        \label{fig:SED}
\end{figure}

The photometric properties, as well as parameters derived from them, are presented in Table \ref{tab:table2}. Gaia DR2 photometry in bands G, G$_{BP}$ and G$_{RP}$, together with interstellar extinctions for the CSPNe from the literature can be used to derive extinction in the G band as well as colour excess E(G$_{BP}$-G$_{RP}$). We assumed that interstellar extinction is the same for the CSPN and the binary companion, and we used \citet{2018A&A...614A..19D} relations to transform the global extinction A$_0$ into the Gaia passbands ones, taking into account the extinction coefficients dependency on colour.  It has to be pointed out that circumstellar extinction is only important for compact PNe which is not the case for any star in our sample. 
 PHR J1129-6012 and PN SB 36 lack  extinction information and, therefore, these two objects have been excluded from our evolutive analysis. The next step was to use PARSEC isochrones in Gaia DR2 passbands (\citealt{2018A&A...616A...4E}) to derive the effective temperature and luminosity values for the companion stars that are listed in Table  \ref{tab:table2}. Note that Abell 33-B and NGC 6853-C do not have colour data in DR2. For these stars, we retrieved the available photometric information from the Spanish Virtual Observatory platform, VOSA (http://svo2.cab.inta-csic.es/theory/vosa/), and we built their spectral energy distribution (SED). Using the VOSA tool, we fitted the SED to NextGen (AGSS2009) models (\citealt{2012RSPTA.370.2765A}) and the result is shown in Fig. \ref{fig:SED}. The values obtained for 
the temperatures and luminosities for these two stars are also listed in Table \ref{tab:table2}. 

\onecolumn


\begin{table}[h]
    \caption{Astrometric data for our comoving systems.}
    \label{tab:table1}
    \begin{tabular}{ l c c c c c c c c c c } 
    
     \hline\hline
     Object & RA & Dec & Parallax & Distance & Separation & $PM_{RA}$ & $PM_{Dec}$ & $V_{Rad}$\\
      & (º) & (º) & ($mas$) & ($pc$) & ($AU$) & ($mas \cdot yr^{-1}$) & ($mas \cdot yr^{-1}$) & ($km \cdot s^{-1}$)\\
     \hline
Abell 24 (CS) & 117.9065 & 3.0059 & $1.46\pm0.15$ & $691_{-59}^{+70}$ & $\cdots$ & $-4.37\pm0.23$ & $-0.75\pm0.14$ & 13 \\	
Abell 24 - B & 117.9067 & 3.0021 & $1.38\pm0.07$ & $725_{-27}^{+29}$ &  9,912 & $-4.25\pm0.11$ & $-0.97\pm0.09$ &$\cdots$
\\
Abell 33 (CS) & 144.788 & -2.8084 & $1.07\pm0.10$ & $932_{-67}^{+77}$ &  $\cdots$& $-14.76\pm0.17$ & $9.42\pm0.15$ & 60
\\
Abell 33 - B* & 144.7878 & -2.8089 & $1.17\pm0.09$ & $856_{-51}^{+57}$ &  1,542 & $-14.94\pm0.15$ & $9.62\pm0.14$ &$\cdots$
\\
Abell 34 (CS) & 146.3973 & -13.1711 & $0.89\pm0.12$ & $1118_{-115}^{+144}$ & $\cdots$& $3.08\pm0.19$ & $-9.13\pm0.24$ &$\cdots$
\\
Abell 34 - B*& 146.3954 & -13.1693 & $0.87\pm0.06$ & $1155_{-51}^{+56}$  & 10,472 & $3.23\pm0.09$ & $-9.11\pm0.11$ &$\cdots$
\\
NGC 246 (CS) & 11.7639 & -11.872 & $1.98\pm0.11$ & $506_{-59}^{+70}$ & $\cdots$& $-16.96\pm0.22$ & $-8.88\pm0.13$ & -16
\\
NGC 246 - B*& 11.7647 & -11.8727 & $1.83\pm0.06$ & $547_{-10}^{+10}$ &  2,118 & $-16.61\pm0.09$ & $-8.76\pm0.08$ & -13
\\
NGC 3699 (CS) & 171.991 & -59.9579 & $0.68\pm0.11$ & $1506_{-205}^{+279}$ & $\cdots$& $-3.19\pm0.16$ & $1.14\pm0.15$ & -16
\\
NGC 3699 - B & 171.9922 & -59.9585 & $0.64\pm0.07$ & $1571_{-123}^{+146}$ & 5,036 & $-3.22\pm0.10$ & $1.07\pm0.10$ &$\cdots$
\\
NGC 6853 (CS) & 299.9016 & 22.7212 & $2.69\pm0.06$ & $372_{-6}^{+6}$ & $\cdots$& $10.39\pm0.09$ & $3.66\pm0.09$ & -42
\\
NGC 6853 - B*& 299.9005 & 22.7197 & $2.62\pm0.06$ & $382_{-6}^{+7}$ &  2,453 & $10.22\pm0.09$ & $3.81\pm0.09$ &$\cdots$
\\
NGC 6853 - C  & 299.8997 & 22.7202 & $2.36\pm0.52$ & $457_{-101}^{+181}$ & 3,322 & $9.13\pm0.64$ & $3.78\pm0.78$ &$\cdots$
\\
PHR J1129-6012 (CS) & 172.4594 & -60.2022 & $0.47\pm0.09$ & $2159_{-320}^{+448}$ & $\cdots$& $-6.64\pm0.14$ & $2.34\pm0.12$ &$\cdots$
\\
PHR J1129-6012 - B & 172.4573 & -60.2022 & $0.41\pm0.09$ & $2482_{-400}^{+576}$ & 9,551 & $-6.84\pm0.38$ & $2.70\pm0.12$ &$\cdots$
\\
PN SB 36 (CS) & 268.5868 & -39.1772 & $0.63\pm0.08$ & $1610_{-156}^{+192}$ & $\cdots$ & $4.43\pm0.11$ & $-4.75\pm0.10$ & 35
\\
PN SB 36 - B & 268.5831 & -39.1761 & $0.77\pm0.08$ & $1331_{-108}^{+128}$ & 14,880 & $4.43\pm0.10$ & $-4.78\pm0.09$ &$\cdots$ \\
     \hline
    \end{tabular}
    
\end{table}    

\tablefoot{Parameters for the CSPNe are taken from 
paper 1, while those for the companion stars and all proper motions are obtained from Gaia DR2. References for the radial velocities of the CSPNe can also be found in 
paper 1. The radial velocity of NGC 246-B was obtained in this work.

(*): Companion star previously identified or suspected, see text for details.}



\small

\begin{table}[h]
    \caption{Photometric and evolutive data for our comoving systems.}
    \label{tab:table2}
    \begin{tabular}{ l c c c c c c c c c } 
    
     \hline\hline
     Object & G & $M_{G}$ & $A_{G}$ & $G_{BP}-G_{RP}$ & $E(G_{BP}-G_{RP})$ & $T_{eff}$ (kK)& $log(L/L_{\odot})$& $M(M_{\odot})$\\
     \hline
Abell 24 (CS)& 17.41 & $7.99$ & 0.22 & $-0.72$ & 0.11 & 137.0 & 1.703 & >3.0 \\
Abell 24 - B& 15.67 & $6.20$ & 0.16 & 0.99 & 0.09 & 5.0 & -0.569 & 0.761 \\
Abell 33 (CS)& 15.96 & $6.12$ & 0.00 & -0.37 & 0.00 & 100.0 & 2.04 & $1.1_{-0.1}^{+0.3}$ \\
Abell 33 - B& 16.67 & $6.89$ &$\cdots$ &$\cdots$&$\cdots$& 4.6 & -0.787 & 0.762 \\
Abell 34 (CS)& 16.42 & $6.04$ & 0.14 & -0.67 & 0.07 & 98.0 & 2.079 & $1.02_{-0.02}^{+0.01}$ \\
Abell 34 - B& 14.85 & $4.43$ & 0.11 & 0.82 & 0.06 & 5.9 & 0.086 & 0.908 \\
NGC 246 (CS)& 11.81 & $3.24$ & 0.05 & -0.68 & 0.03 & 150 & 3.688 & 3.75 \\
NGC 246 - B& 14.19 & $5.46$ & 0.38 & 0.91 & 0.02 & 5.5 & -0.314 & 0.846 \\
NGC 3699 (CS)& 17.60 & $4.93$ & 1.78 & -0.55 & 0.38 &$\cdots$&$\cdots$&$\cdots$\\
NGC 3699 - B& 16.54 & $4.02$ & 1.54 & 0.40 & 0.70 & 6.5 & 0.274 & 1.161\\
NGC 6853 (CS)& 14.03 & $6.02$ & 0.15 & -0.67 & 0.08 & 135.0 & 2.44 & $2.0_{-0.4}^{+0.4}$ \\
NGC 6853 - B& 16.12 & $8.10$ & 0.11 & 1.24 & 0.06 & 4.0 & -1.119 & 0.587 \\
NGC 6853 - C& $18.99$ & $10.80$ &$\cdots$&$\cdots$&$\cdots$& 3.3 & -2.020 & 0.200 \\
PHR J1129-6012 (CS)& $17.26$ & $5.33^{*}$ &$\cdots$& $1.06^{*}$ &$\cdots$&$\cdots$&$\cdots$&$\cdots$\\
PHR J1129-6012 - B & $17.12$ & $4.83^{*}$ &$\cdots$& $1.06^{*}$ &$\cdots$&$\cdots$&$\cdots$&$\cdots$&\\
PN SB 36 (CS)& 15.25 & $4.03^{*}$ &$\cdots$& $1.23^{*}$ &$\cdots$&$\cdots$&$\cdots$&$\cdots$\\
PN SB 36 - B& 14.62 & $3.85^{*}$ & $\cdots$ & $0.93^{*}$ &$\cdots$& $\cdots$&$\cdots$&$\cdots$\\
     \hline
    \end{tabular}
    
\end{table}    

\tablefoot{Magnitudes are obtained from Gaia DR2,
while extinctions are derived by the method of \citet{2018A&A...614A..19D} using interestellar extinctions taken from paper 1. Temperatures and luminosities are taken from  paper 1 in the case of the CSPNe, while in the case of the companion stars they are estimated either from PARSEC isochrones in Gaia DR2 passbands, \citet{2018A&A...616A...4E}; or by fitting the corresponding SED to NextGen (AGSS2009) models using VOSA platform (http://svo2.cab.inta-csic.es/theory/vosa/index.php), \citet{2012RSPTA.370.2765A}. Masses for the CSPNe are for the progenitor stars, and masses for the comoving companions were calculated either from the PARSEC models or from the SED fits. See text for details. 

(*): Value not corrected from extinction.}

\twocolumn
\normalsize

Information in Table \ref{tab:table2} can be used to explore the evolutionary properties of the companion stars. All companions with DR2 photometry show positions close to the Zero Age Main Sequence (ZAMS), except Abell 34-B that has already evolved into the subgiant branch. PARSEC evolutionary tracks were interpolated to derive the mass values that are listed in the last column of Table \ref{tab:table2}. Taking into account the low errors in DR2 photometry (which have a complex dependence on the number of visits of the satellite and on the brightness of the sources, among other variables), we estimate that our masses have uncertainties below $20\%$. By assuming the values obtained for the masses and the positions of the objects on the Main Sequence, we can also infer the spectral types of these stars.  
The next step was to compare the mass values obtained for the companion stars with those of the CSPNe. In the case of Abell 33, Abell 34, and NGC 6853 central stars, we adopted the temperature and luminosity values published in 
paper 1.
We have not found a reliable effective temperature measurement published for \hbox{NGC 3699} central star and so this system was excluded from the evolutionary analysis. In \citet{1989ApJ...345..871K}, they estimated that such temperature must be high, of the order of 260 kK. But using the mentioned value, the bolometric correction would lead to an unrealistically high luminosity for this object. The evolutionary parameters of Abell 24 CS are also subject to high uncertainty. Its location in the HR diagram as calculated in 
paper 1 places it at a rather high temperature for its luminosity, and by interpolating \citet{2016A&A...588A..25M} evolutionary tracks only a lower limit of 3 $M_{\odot}$ for the progenitor star could be obtained. We would like to call 
attention on the fact that by using a lower extinction value for this object, of A$_{V}$ = 0.055, as obtained using the DR2 distance and Bayestar19 3D Dust Maps \citep{2019ApJ...887...93G}, a position in the HR diagram is obtained which is compatible with the \citet{2016A&A...588A..25M} evolutionary track for a progenitor star of 3 $M_{\odot}$. 

NGC 246 central star is a H-deficient CSPN, \citet{1975ApJ...196..195H}. It is an extremely hot member of the PG1159 spectroscopic subtype, within which it is the brightest (m$_{V}$ = 11.78) star. \citet{2006ASPC..348..191W} fitted NLTE stellar atmosphere models to FUSE and HST/STIS spectra obtaining Teff = 130 kK and log g = 5.7. Using post-AGB evolutionary models calculated under a born-again scenario from a very late thermal pulse (VLTP) by \citet{2006A&A...454..845M}, they derived a stellar mass of M = 0.74 $M_{\odot}$ which would be the remnant of a Main Sequence star of approximately 3.75 $M_{\odot}$ \citep{2009ApJ...704.1605A}.

Evolutive masses for H-rich post-AGB stars can be derived by fitting evolutionary tracks by \citet{2016A&A...588A..25M} and the progenitor mass obtained from their self-consistent models that include all previous evolutionary stages from the ZAMS to the white dwarf phase. Table \ref{tab:table2} shows the progenitor masses derived for the CSPNe together with low and high uncertainty values. These uncertainties are estimated from the luminosity and temperature errors in the HR diagram derived 
in paper 1. In the following sections we shall comment on the results obtained for the individual binary pairs. 

\subsection{Abell 24}
Abell 24 (Fig. \ref{fig:images}) is an elliptical PN located 691 pc away from the Sun. It is a large nebula with a mean radius of 0.633 pc ($\sim$130,000 AU). We have found a wide binary companion, Abell 24-B, with a projected separation of 9,912 AU from the CS, and by fitting its photometry to PARSEC evolutionary tracks we obtained a mass of 0.76 $M_{\odot}$ which corresponds to an early K-type star.  Meanwhile, a lower-limit mass of 3 $M_{\odot}$ can be derived for Abell 24 CS from comparison with \citet{2016A&A...588A..25M} post-AGB evolutionary tracks. 
\citet{Ciardullo_1999} commented on the presence of a possible binary pair in this PN, but they identified a different companion, closer to the CSPN, that 
finally was classified as a "doubtful association". If we analyse this other candidate in Gaia DR2, we find that neither parallax, nor proper motions are compatible with those from the CS, so, we can discard a binary bound in this case.

\subsection{Abell 33}

This PN (Fig. \ref{fig:images}) has a round shape and it is located 932 pc away from the Sun. It has a medium radius of 0.516 pc ($\sim$106,000 AU). We have detected a wide binary companion with a projected separation of 1,542 AU from the CS. This star was already mentioned as a possible binary companion by \citet{Ciardullo_1999}, and this is now confirmed by DR2 astrometry. 

Using the VOSA tool we fitted the SED of Abell 33-B to NextGen (AGSS2009) models to  and we obtained a value of 0.76 $M_{\odot}$, which corresponds to an early K-type star, while a value of 1.13 $M_{\odot}$ was derived for Abell 33 CS from postAGB tracks in \citet{2016A&A...588A..25M}. 

\subsection{Abell 34}

This PN (Fig. \ref{fig:images}) is located 
1,118 pc away from the Sun. It has a round shape and a large radius of 0.778 pc ($\sim$160,000 AU). We have found a comoving star with a projected separation of 10,472 AU from the CS. \citet{Ciardullo_1999} already mentioned the existence of this possible binary companion, although they calculated a probability of 28\% for it to correspond to a chance alignment. Gaia DR2 astrometry allows us to confirm that this star is a true wide binary companion of the CSPN. 

\begin{figure}[h!]
        \includegraphics[width=9cm, height=6.5cm]{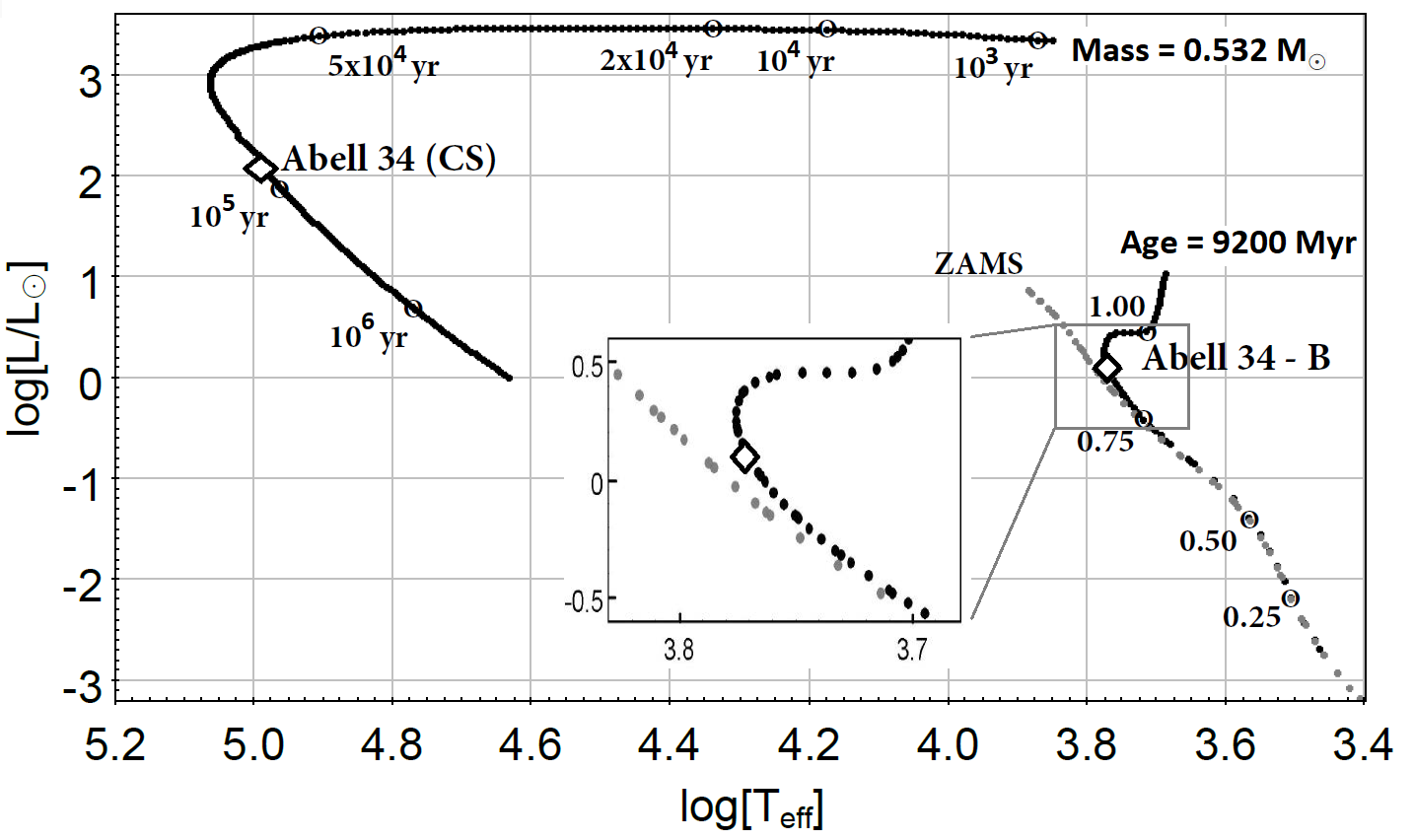}
        \caption{HR diagram showing the position of both Abell 34 CS and its comoving companion star (rhomboid symbols). The CS is plotted together with the 0.532 $M_{\odot}$ evolutionary track from \citet{2016A&A...588A..25M}, that corresponds to a mass of 1.0 $M_{\odot}$ in the MS. Ages shown on the track refer to the post-AGB phase. In order to obtain the age from the MS, a value of 9244 Myr must be added. The companion star is plotted together with the 9200 Myr isochrone and ZAMS from PARSEC models in Gaia passbands (\citealt{2018A&A...616A...4E}).
        Masses are expressed in units of solar mass ($M_{\odot}$).}
        \label{fig:abell_34}
\end{figure}

Abell 34-B has already evolved from the MS, it is a subgiant star, and by fitting its photometry to PARSEC evolutionary tracks we obtained a mass of 0.91 $M_{\odot}$, a spectral type G2-5IV and an evolutionary age of 9,200 Myrs. If we fit the models for postAGB stars from \citet{2016A&A...588A..25M},
we obtain a MS mass of 1.02$M_{\odot}$ and an age of 9,244 Myrs for Abell 34 CS. These values allow us to confirm a common binary scenario for the evolution of these stars. An HR diagram showing the position of both Abell 34 CS and its comoving companion star is shown in Fig. \ref{fig:abell_34}.


\subsection{NGC 246}

This PN (Fig. \ref{fig:images}), also known as Skull nebula, is located at a distance of 506 pc from the Sun. It has an elliptical morphology and a medium size, with a mean radius of 0.298 pc ($\sim$61,000 AU). We have detected a comoving star with a projected separation of 2,118 AU from the CSPN. According to \citet{10.1093/mnras/stu1677}, this PN has associated a triple system of stars, with two companions with separations around 500 AU and 2,000 AU. In such a context, we have detected the farthest one, while the closest one cannot be resolved in the light of Gaia DR2 data. From the photometry fitting to PARSEC models, NGC 246-B has a mass of 0.85 $M_{\odot}$ and a spectral type G8-5, while we assume a progenitor mass for NGC 246 CS of 3.75 $M_{\odot}$ as previously discussed. An optical spectrum recently obtained at the INT telescope using the IDS spectrograph allows us to confirm an intermediate spectral type for NGC 246-B, as well as to obtain a radial velocity value of $-13 \pm8$ $km \cdot s^{-1}$, compatible with the known value for the central star of -16 $km \cdot s^{-1}$.

\subsection{NGC 3699}
NGC 3699 (Fig. \ref{fig:images}) is located at a distance of 1,506 pc  from the Sun.
It has a bipolar morphology, 
a small mean radius of 0.153 pc ($\sim$32,000 AU) and presents a quite high extinction in G passband. We have identified a binary companion with a projected separation of 5,036 AU from the CS. By fitting DR2 photometry to PARSEC models to  we obtained a mass of 1.16 $M_{\odot}$ and late-F spectral type for this star.

\subsection{NGC 6853}
The NGC 6853 nebula (Fig. \ref{fig:images}), also known as Dumbbel Nebula, is located at only 372 pc from the Sun. It has a bipolar shape and a medium mean radius of 0.368 pc ($\sim$76,000 AU). In this case, we have detected two wide binary companions, one of them with a projected separation of 2,453 AU from the CS and the other with a projected separation of 3,322 AU. For the former,
DR2 photometry allowed to derive a mass of 0.59 $M_{\odot}$ and a late-K spectral type. The latter comoving star,
is faint (G = 18.99) and Gaia colour (G$_{BP}$-G$_{RP}$) is not available for it. By fitting the SED available at the SVO with the VOSA tool we found that it is a late M-type star with a mass value of 0.2 $M_{\odot}$.  NGC 6853 is included in the David Jones catalogue (http://www.drdjones.net/bcspn/) of binary PNe. The existence of a binary in NGC 6853 was first suggested by \citet{1973PASP...85..401C}, who identified a companion with V = 17 located at 6.5 arcsecs from the central star, which is coincidental with the brightest of our wide companions. In 1977, astrometric measurements by Cudworth confirmed that the proper motions agree with those of the CS. These results are further confirmed by Gaia DR2 astrometry. Regarding NGC 6853-C, it is the faintest star in our sample, with the highest errors in the astrometric parameters, 22\% in parallax, but with overall values that meet our selection criteria. Forthcoming EDR3 astrometry will help to clarify if B and C stars form a bound pair orbiting  NGC 6853 CS.



\subsection{PHR J1129-6012}
PHR J1129-6012 (Fig. \ref{fig:images}) is an elliptical PN located quite far from the Sun, at 2,159 pc. It is a small nebula, with a mean radius of 0.099 pc ($\sim$20,000 AU). We have detected a comoving object with a projected separation of 9,551 AU from the CS. Both stars in the system have similar brightness (mag $G \approx 17$) and similar Gaia colour ($G_{BP}-G_{RP} \approx 1$) which seems interesting and deserves further study. Note that a high interstellar extinction is expected for this nebula, located at a very low galactic latitude, 
this fact perhaps explains the red colour observed for the CS. 
Very recently, \citet{2020A&A...638A.103C}, have highlighted the fact that CSPN identifications in commonly used PN catalogues are highly uncertain, and they published a new catalogue of CSPN using DR2 astrometry and colours ($G_{BP}-G_{RP}$) to more accurately identify the bluest and more centred star within a nebula. In their catalogue they assign a probability of being the CSPN of only 40\% to PHR J1129-6012 CS. In fact, HASH database provides a spectrum of this source that allows to derive an extinction value E(B-V) close to 1. In that case, it would imply that the absolute brightness of this pair of stars would not correspond to that of a CSPN and therefore the identification would not be correct.

\subsection{PN SB 36}
PN SB 36 (Fig. \ref{fig:images}) is also a very small PN, with a mean radius of 0.029 pc ($\sim$6,000 AU). It is
located 1,610 pc away from the Sun, close to the galactic plane. We have identified a comoving object with a projected separation of 14,880 AU from the CS. Although PN SB 36 is included in our GAPN sample (concretely from \citealt{kerber03}), there are doubts about if it is really a PN, as its spectrum in HASH database shows only $H_{\alpha}$ emission.








\section{Conclusions}
By using Gaia DR2 astrometry (parallaxes and proper motions), we have found 9 wide binary companions associated to 8 CSPNe in the GAPN sample (Golden Astrometry Planetary Nebulae, \citealt{2019A&A...630A.150G}),
with projected separations of less than 15,000 AU in all cases. When we searched for comoving objects to these CSPNe, we applied quite strict selection criteria in order to avoid 
any false comoving object detection. Furthermore, we did a thorough analysis to estimate the probability of having detected by chance any of the companion stars, obtaining very low values. 

Some of the binary pairs were already listed as possible binaries in the literature, and Gaia DR2 astrometry has allowed us to confirm them as such. {We also found two of the CS identifications as doubtful, although the binary pairs are potentially related.} Our procedure allows to identify a limited number of 
systems due to the errors present in DR2 astrometry, but it 
is also a useful introductory methodology potentially allowing to identify larger numbers of binary systems in the future, when Gaia DR3 and later archives become available.
By using appropriate evolutionary models, we could estimate masses for both the CS and the binary companions, that are compatible with a joint evolution scenario.
Even more, the companion of Abell 34 CS has already evolved out of the MS, and we were able to obtain a value for its age, which satisfactorily agrees with the age obtained for the nebula central star.


In this work we have limited ourselves to the search for distant binary partners. In a recent study, \citet{2017ApJ...837L..10B} propose that hierarchical triple systems can explain
the morphologies of PNe, displaying large departures from axisymmetrical and point-symmetrical structures, and in some cases from mirror-symmetry. Among our eight listed nebulae we find a variety of morphologies, such as round (Abell 33, Abell 34 and PN SB 36), elliptical (Abell 24, NGC 246 and PHR J1129-6012), and bipolar (NGC 3699 and NGC 6853), so there is no correlation between wide binaries and the nebular morphology.
However, multi-band, high sensitivity, and high spatial resolution images would be needed to distinguish morphological structures that could give us clues on the possible influence such wide companions might have had.

\citet{2017ApJ...837L..10B} pointed out that very wide tertiary stars, or even higher-order hierarchical systems at
hundreds of AU, cannot play a significant role in shaping the morphologies. All our wide binary companions are too far from the central star to influence their nebula morphology under the mechanisms proposed by these authors. Nevertheless, one of our objects,  NGC 3699, is included in their catalogue under the category of objects "shaped by a triple system". We will have to wait for the forthcoming Gaia DR3 release to expand the search for binaries to objects located closer to the CS to be able to contribute with greater statistics to the study of the influence of binary systems, hierarchical or not, in the shaping of PNe.



\begin{acknowledgements}
        This work has made use of data from the European Space Agency (ESA) Gaia mission and processed by the Gaia Data Processing and Analysis Consortium (DPAC). Funding for the DPAC has been provided by national institutions, in particular the institutions participating in the Gaia Multilateral Agreement. This research has made use of the SIMBAD database, operated at CDS, Strasbourg, France, and the ALADIN applet. The authors have also made use of VOSA, developed under the Spanish Virtual Observatory project supported by the Spanish MINECO through grant AyA2017-84089, and partially updated thanks to the EU Horizon 2020 Research and Innovation Programme, under Grant 776403 (EXOPLANETS-A). The INT is operated on the island of La Palma by the Isaac Newton Group of Telescopes in the Spanish Observatorio del Roque de los Muchachos of the Instituto de Astrofísica de Canarias. Funding from Spanish Ministry projects ESP2016-80079-C2-2-R, RTI2018-095076-B-C22, Xunta de Galicia ED431B 2018/42, and  AYA-2017-88254-P is acknowledged by the authors. IGS acknowledges financial support from the Spanish National Programme for the Promotion of Talent and its Employability grant BES-2017-083126 cofunded by the European Social Fund. We also wish to acknowledge the support received from the  CITIC, funded by Xunta de Galicia and the European Union (FEDER Galicia 2014-2020 Program) by grant ED431G 2019/01.
\end{acknowledgements}

\bibliographystyle{aa} 

\bibliography{bibliopn}

\begin{thebibliography}{30}
\expandafter\ifx\csname natexlab\endcsname\relax\def\natexlab#1{#1}\fi

\bibitem[{Adam \& Mugrauer(2014)}]{10.1093/mnras/stu1677}
Adam, C. \& Mugrauer, M. 2014, \mnras, 444, 3459

\bibitem[{{Allard} {et~al.}(2012){Allard}, {Homeier}, \&
  {Freytag}}]{2012RSPTA.370.2765A}
{Allard}, F., {Homeier}, D., \& {Freytag}, B. 2012, Philosophical Transactions
  of the Royal Society of London Series A, 370, 2765

\bibitem[{{Althaus} {et~al.}(2009){Althaus}, {Panei}, {Miller Bertolami},
  {Garc{\'\i}a-Berro}, {C{\'o}rsico}, {Romero}, {Kepler}, \&
  {Rohrmann}}]{2009ApJ...704.1605A}
{Althaus}, L.~G., {Panei}, J.~A., {Miller Bertolami}, M.~M., {et~al.} 2009,
  \apj, 704, 1605

\bibitem[{{Bayo} {et~al.}(2008){Bayo}, {Rodrigo}, {Barrado y Navascu{\'e}s},
  {Solano}, {Guti{\'e}rrez}, {Morales-Calder{\'o}n}, \&
  {Allard}}]{2008A&A...492..277B}
{Bayo}, A., {Rodrigo}, C., {Barrado y Navascu{\'e}s}, D., {et~al.} 2008, \aap,
  492, 277

\bibitem[{{Bear} \& {Soker}(2017)}]{2017ApJ...837L..10B}
{Bear}, E. \& {Soker}, N. 2017, \apjl, 837, L10

\bibitem[{{Boffin} \& {Jones}(2019)}]{2019ibfe.book.....B}
{Boffin}, H. M.~J. \& {Jones}, D. 2019, {The Importance of Binaries in the
  Formation and Evolution of Planetary Nebulae}, ed. M.~{Ratcliffe}, I.~M.
  {Hillebrandt}, Wolfgagng, \& D.~{Weintraub}, SpringerBriefs in Astronomy

\bibitem[{{Chornay} \& {Walton}(2020)}]{2020A&A...638A.103C}
{Chornay}, N. \& {Walton}, N.~A. 2020, \aap, 638, A103

\bibitem[{Ciardullo {et~al.}(1999)Ciardullo, Bond, Sipior, Fullton, Zhang, \&
  Schaefer}]{Ciardullo_1999}
Ciardullo, R., Bond, H.~E., Sipior, M.~S., {et~al.} 1999, \aj, 118, 488–508

\bibitem[{{Close} {et~al.}(1990){Close}, {Richer}, \&
  {Crabtree}}]{1990AJ....100.1968C}
{Close}, L.~M., {Richer}, H.~B., \& {Crabtree}, D.~R. 1990, \aj, 100, 1968

\bibitem[{{Cudworth}(1973)}]{1973PASP...85..401C}
{Cudworth}, K.~M. 1973, \pasp, 85, 401

\bibitem[{{Danielski} {et~al.}(2018){Danielski}, {Babusiaux}, {Ruiz-Dern},
  {Sartoretti}, \& {Arenou}}]{2018A&A...614A..19D}
{Danielski}, C., {Babusiaux}, C., {Ruiz-Dern}, L., {Sartoretti}, P., \&
  {Arenou}, F. 2018, \aap, 614, A19

\bibitem[{{Evans} {et~al.}(2018){Evans}, {Riello}, {De Angeli}, {Carrasco},
  {Montegriffo}, {Fabricius}, {Jordi}, {Palaversa}, {Diener}, {Busso},
  {Cacciari}, {van Leeuwen}, {Burgess}, {Davidson}, {Harrison}, {Hodgkin},
  {Pancino}, {Richards}, {Altavilla}, {Balaguer-N{\'u}{\~n}ez}, {Barstow},
  {Bellazzini}, {Brown}, {Castellani}, {Cocozza}, {De Luise}, {Delgado},
  {Ducourant}, {Galleti}, {Gilmore}, {Giuffrida}, {Holl}, {Kewley}, {Koposov},
  {Marinoni}, {Marrese}, {Osborne}, {Piersimoni}, {Portell}, {Pulone},
  {Ragaini}, {Sanna}, {Terrett}, {Walton}, {Wevers}, \&
  {Wyrzykowski}}]{2018A&A...616A...4E}
{Evans}, D.~W., {Riello}, M., {De Angeli}, F., {et~al.} 2018, \aap, 616, A4

\bibitem[{{Garcia-Segura}(1997)}]{GarciaSegura97}
{Garcia-Segura}, G. 1997, \apjl, 489, L189

\bibitem[{{Gonz{\'a}lez-Santamar{\'\i}a}
  {et~al.}(2019){Gonz{\'a}lez-Santamar{\'\i}a}, {Manteiga}, {Manchado}, {Ulla},
  \& {Dafonte}}]{2019A&A...630A.150G}
{Gonz{\'a}lez-Santamar{\'\i}a}, I., {Manteiga}, M., {Manchado}, A., {Ulla}, A.,
  \& {Dafonte}, C. 2019, \aap, 630, A150

\bibitem[{{Green} {et~al.}(2019){Green}, {Schlafly}, {Zucker}, {Speagle}, \&
  {Finkbeiner}}]{2019ApJ...887...93G}
{Green}, G.~M., {Schlafly}, E., {Zucker}, C., {Speagle}, J.~S., \&
  {Finkbeiner}, D. 2019, \apj, 887, 93

\bibitem[{{Heap}(1975)}]{1975ApJ...196..195H}
{Heap}, S.~R. 1975, \apj, 196, 195

\bibitem[{{Jim{\'e}nez-Esteban} {et~al.}(2019){Jim{\'e}nez-Esteban}, {Solano},
  \& {Rodrigo}}]{2019AJ....157...78J}
{Jim{\'e}nez-Esteban}, F.~M., {Solano}, E., \& {Rodrigo}, C. 2019, \aj, 157, 78

\bibitem[{{Jones}(2020)}]{2020Galax...8...28J}
{Jones}, D. 2020, Galaxies, 8, 28

\bibitem[{{Jones} \& {Boffin}(2017)}]{2017NatAs...1E.117J}
{Jones}, D. \& {Boffin}, H. M.~J. 2017, Nature Astronomy, 1, 0117

\bibitem[{{Kaler} \& {Jacoby}(1989)}]{1989ApJ...345..871K}
{Kaler}, J.~B. \& {Jacoby}, G.~H. 1989, \apj, 345, 871

\bibitem[{{Kerber} {et~al.}(2003){Kerber}, {Mignani}, {Guglielmetti}, \&
  {Wicenec}}]{kerber03}
{Kerber}, F., {Mignani}, R.~P., {Guglielmetti}, F., \& {Wicenec}, A. 2003,
  \aap, 408, 1029

\bibitem[{{Lindegren} {et~al.}(2018){Lindegren}, {Hern{\'a}ndez}, {Bombrun},
  {Klioner}, {Bastian}, {Ramos-Lerate}, {de Torres}, {Steidelm{\"u}ller},
  {Stephenson}, {Hobbs}, {Lammers}, {Biermann}, {Geyer}, {Hilger}, {Michalik},
  {Stampa}, {McMillan}, {Casta{\~n}eda}, {Clotet}, {Comoretto}, {Davidson},
  {Fabricius}, {Gracia}, {Hambly}, {Hutton}, {Mora}, {Portell}, {van Leeuwen},
  {Abbas}, {Abreu}, {Altmann}, {Andrei}, {Anglada}, {Balaguer-N{\'u}{\~n}ez},
  {Barache}, {Becciani}, {Bertone}, {Bianchi}, {Bouquillon}, {Bourda},
  {Br{\"u}semeister}, {Bucciarelli}, {Busonero}, {Buzzi}, {Cancelliere},
  {Carlucci}, {Charlot}, {Cheek}, {Crosta}, {Crowley}, {de Bruijne}, {de
  Felice}, {Drimmel}, {Esquej}, {Fienga}, {Fraile}, {Gai}, {Garralda},
  {Gonz{\'a}lez-Vidal}, {Guerra}, {Hauser}, {Hofmann}, {Holl}, {Jordan},
  {Lattanzi}, {Lenhardt}, {Liao}, {Licata}, {Lister}, {L{\"o}ffler},
  {Marchant}, {Martin-Fleitas}, {Messineo}, {Mignard}, {Morbidelli}, {Poggio},
  {Riva}, {Rowell}, {Salguero}, {Sarasso}, {Sciacca}, {Siddiqui}, {Smart},
  {Spagna}, {Steele}, {Taris}, {Torra}, {van Elteren}, {van Reeven}, \&
  {Vecchiato}}]{lindegren18}
{Lindegren}, L., {Hern{\'a}ndez}, J., {Bombrun}, A., {et~al.} 2018, \aap, 616,
  A2

\bibitem[{{Manchado}(2004)}]{2004ASPC..313....3M}
{Manchado}, A. 2004, Astronomical Society of the Pacific Conference Series,
  Vol. 313, {Correlation of PN Morphologies and Nebular Parameters}, ed.
  M.~{Meixner}, J.~H. {Kastner}, B.~{Balick}, \& N.~{Soker}, 3

\bibitem[{{Miller Bertolami}(2016)}]{2016A&A...588A..25M}
{Miller Bertolami}, M.~M. 2016, \aap, 588, A25

\bibitem[{{Miller Bertolami} \& {Althaus}(2006)}]{2006A&A...454..845M}
{Miller Bertolami}, M.~M. \& {Althaus}, L.~G. 2006, \aap, 454, 845

\bibitem[{{Sahai} \& {Trauger}(1998)}]{1998AJ....116.1357S}
{Sahai}, R. \& {Trauger}, J.~T. 1998, \aj, 116, 1357

\bibitem[{{Soker}(2001)}]{2001ApJ...558..157S}
{Soker}, N. 2001, \apj, 558, 157

\bibitem[{{Weinberg} {et~al.}(1987){Weinberg}, {Shapiro}, \&
  {Wasserman}}]{1987ApJ...312..367W}
{Weinberg}, M.~D., {Shapiro}, S.~L., \& {Wasserman}, I. 1987, \apj, 312, 367

\bibitem[{{Werner} {et~al.}(2006){Werner}, {Rauch}, \&
  {Kruk}}]{2006ASPC..348..191W}
{Werner}, K., {Rauch}, T., \& {Kruk}, J.~W. 2006, in Astronomical Society of
  the Pacific Conference Series, Vol. 348, Astrophysics in the Far Ultraviolet:
  Five Years of Discovery with FUSE, ed. G.~{Sonneborn}, H.~W. {Moos}, \& B.~G.
  {Andersson}, 191

\bibitem[{{Zavada} \& {P{\'\i}{\v{s}}ka}(2020)}]{2020AJ....159...33Z}
{Zavada}, P. \& {P{\'\i}{\v{s}}ka}, K. 2020, \aj, 159, 33

\end{thebibliography}


\begin{appendix}

\onecolumn


%
    

\section{Images}

\begin{figure}[h!]
        \includegraphics[width=6.1cm,height=6cm]{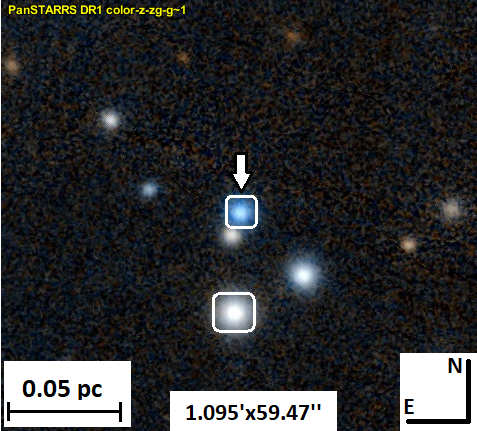}
        \includegraphics[width=6.1cm,height=6cm]{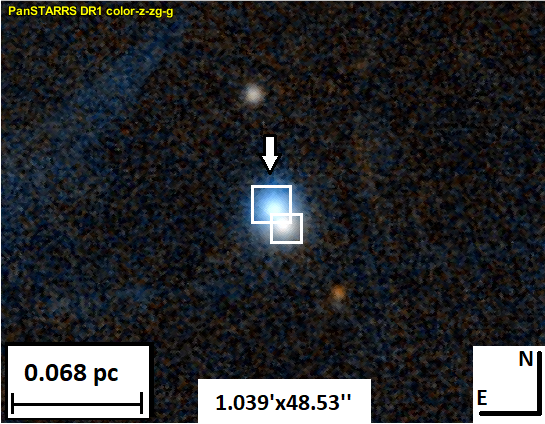}
        \includegraphics[width=6.1cm,height=6cm]{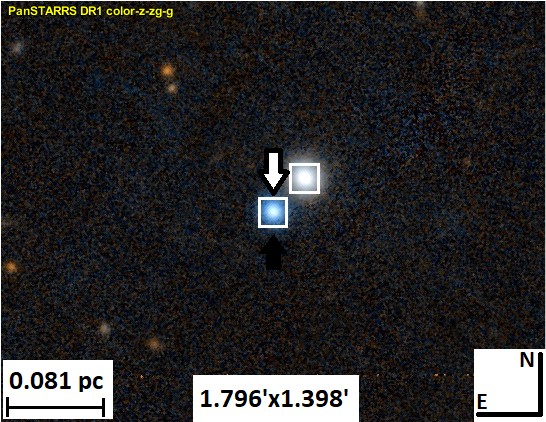}
        \includegraphics[width=6.1cm,height=6cm]{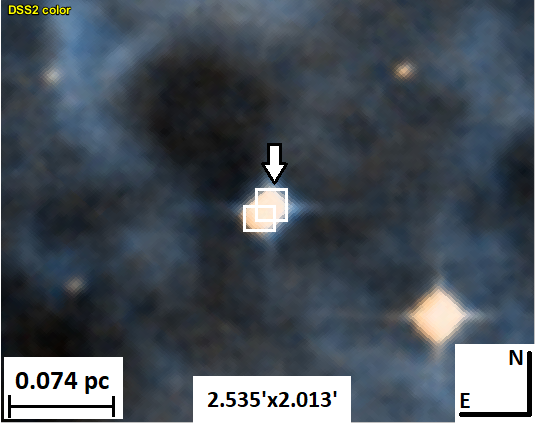}
        \includegraphics[width=6.1cm,height=6cm]{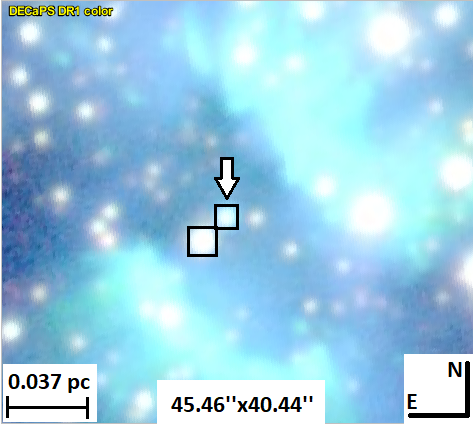}
        \includegraphics[width=6.1cm,height=6cm]{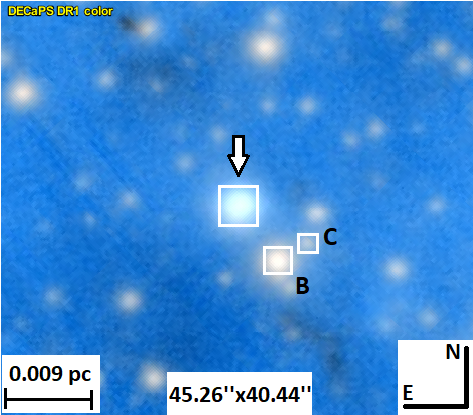}
        \includegraphics[width=6.1cm,height=6cm]{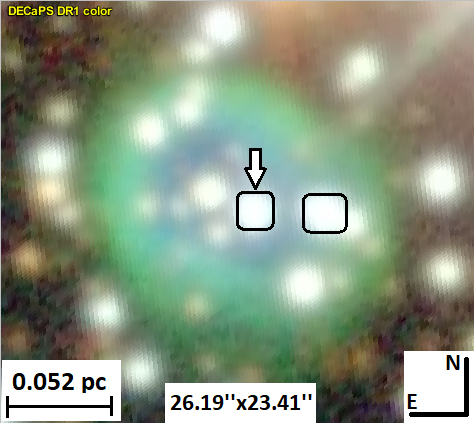}
        \includegraphics[width=6.1cm,height=6cm]{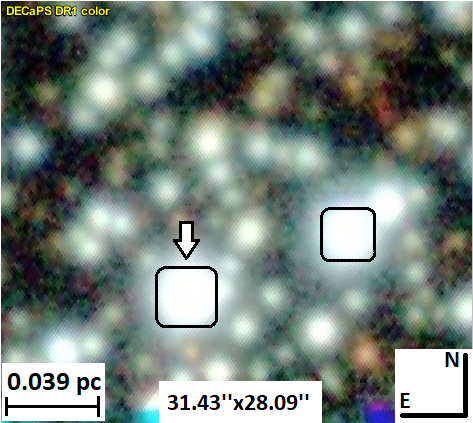}
        \caption{Identification images of Abell 24, Abell 33, Abell 34, NGC 246, NGC 3699, NGC 6853, PHR J1129-6012 and PN SB 36 nebulae (from left to right and from up to down), showing the location of both the CSPN (arrow) and the comoving companions. Images from ALADIN applet.}
        \label{fig:images}
\end{figure}

\clearpage

\end{appendix}

\end{document}